\documentclass[11pt,a4paper]{article}   
\usepackage{amsmath}
\usepackage{amssymb}
\usepackage{mathrsfs}

\usepackage{rotating}

\usepackage[round]{natbib}

\newcommand{\wrt}{ ~ {\rm d}}

\renewcommand{\exp}{{\rm e}}

\newcommand{\eps}{\varepsilon}

\newcommand{\ds}{\displaystyle}

\newcommand{\eg}{e.g.\ }
\newcommand{\ie}{i.e.\ }

\newcommand{\str}{E}

\newcommand{\diss}{\alpha}
\newcommand{\dissn}{a}
\newcommand{\pdt}{p}
\newcommand{\fdm}{d}
\newcommand{\dmx}{d_{\rm mx}}
\newcommand{\dmn}{d_{\rm mn}}
\newcommand{\dav}{\langle d\rangle}

\begin{document}
\title{An idealised wave-ice interaction model without subgrid spatial and temporal discretisations.}
\author{L.~G.~Bennetts$^{1}$, S.~O'Farrell$^{2}$, P.~Uotila$^{2,3}$ and V.~A.~Squire$^{4}$
\\
{\footnotesize
$^{1}$School of Mathematical Sciences, University of Adelaide, Australia}
\\
{\footnotesize
$^{2}$\textsc{Csiro} Marine and Atmospheric Research, Aspendale, Australia}
\\
{\footnotesize
$^3$Finnish Meteorological Institute, Helsinki, Finland}
\\
{\footnotesize
$^{4}$Department of Mathematics and Statistics, University of Otago, New Zealand}
}
\date{\today}
\maketitle

\begin{abstract}
A modified version of the wave-ice interaction model proposed by 
\citet{Wiletal13a,Wiletal13b} is presented for an idealised transect geometry.
Wave attenuation due to ice floes and 
wave-induced ice fracture are both included in the wave-ice interaction model.
Subgrid spatial and temporal discretisations are not required in the
modified version of the model, thereby facilitating its future 
integration into large-scaled coupled models.
Results produced by the new model are compared to results produced by the original model of 
\citet{Wiletal13b}.
\end{abstract}


\section*{Introduction}

Ocean surface waves penetrate tens to hundreds of kilometres into the sea ice-covered ocean, 
causing the ice to bend and flex.
The ice cover can only endure a certain degree of flexure before it fractures into floes with diameters on the order of the local wavelengths.
In this way, waves regulate maximum floe sizes.
Simultaneously, the presence of the ice-cover attenuates the waves.
Waves therefore only remain strong enough to fracture the ice for a finite distance.
In this work, the region of the ice-covered ocean in which floe sizes are controlled by wave-induced fracture is used to define the marginal ice zone (\textsc{miz}).
Moreover, ice-cover filters the wave spectrum preferentially towards low frequencies, 
so that wavelengths and, consequently, maximum floe sizes generally increase with distance into the \textsc{miz}.

Wave attenuation in the \textsc{miz} is modelled as an accumulation of partial wave reflections and transmissions by floes \citep[see][and references therein]{Ben&Squ12a}.
Attenuation models based on viscous effects have also been proposed \citep[e.g.][]{Wan&She10}.
Comparatively few fracture models exist.
The fracture models that do exist are based on wave-induced
flexural motion of the ice resulting in strains that exceed a threshold, dependent on ice thickness and rigidity.

Until recently wave attenuation and ice fracture models have been independent from one another, 
\ie attenuation models consider the floe size distribution to be known, and ice fracture models consider the distribution of wave energy to be known.
However, wave attenuation and ice fracture are coupled processes. 
The rate of wave attenuation depends on the floe size distribution, which is controlled by wave-induced fracture.
Ice fracture depends on the local wave energy, which is controlled by wave attenuation imposed by ice cover.

The wave-ice interaction model (\textsc{wim}) developed by \citet{Dumetal11} and \citet{Wiletal13a,Wiletal13b} 
is the first to model wave attenuation and ice fracture as coupled processes. 
The \textsc{wim} predicts the floe size distribution and the distribution of wave energy in the \textsc{miz} simultaneously, 
given an incident wave forcing from the open ocean and properties of the ice cover.

\citet{Dumetal11} and \citet{Wiletal13a,Wiletal13b} designed the \textsc{wim} as a link between wave and sea ice model components of large-scale operational ice-ocean forecasting models.
The \textsc{wim} is nested in regions of operational interest in the large-scale models using refined spatial and temporal grids.
The \textsc{wim} then provides prognostic information regarding the \textsc{miz} floe size distribution and wave activity in the ice-covered ocean, 
thus allowing for more accurate safety forecasts for offshore engineering activities.

More generally, the \textsc{wim} provides an opportunity to integrate prognostic  \textsc{miz} wave and floe size information  
into oceanic general circulation models (\textsc{ogcm}s), used for \eg climate studies, and here considered as containing wave and sea ice model components.
The information will improve the accuracy of existing floe size dependent processes in 
\textsc{ogcm}s, \eg form drag \citep{Tsaetal_inpress},
and promote the development of new models of wave and floe size dependent processes, 
\eg accelerated melt of floes due to wave overwash \citep{Mas&Sta10}.

Subgridding incurs a large computational cost, and is therefore not a viable option to implement 
the \textsc{wim} in an \textsc{ogcm} on a circumpolar or global scale.
An alternative numerical implementation of the \textsc{wim} is proposed here.
The new model does not involve subgrid spatial or temporal discretisations.
Instead, the floe size distribution and the rate of attenuation are balanced in the cell.
An idealised version of the new model is presented, in which
a single transect of the ice-covered ocean is considered.
Results produced by the new model are compared with results produced by the model of \citet{Wiletal13b}.


\section*{Model description}

Consider a transect of the ocean surface within or containing a region of the \textsc{miz}.
Points on the transect are defined by the coordinate $x\in(0,L)$.
The transect represents a single cell in an \textsc{ogcm}. 
The time frame under consideration, $t\in(0,T)$, 
is equivalent to a single time step in the \textsc{ogcm}.
For climate models $L\sim 50$\,km and $T\sim 1$\,h.
The sea ice model component of the \textsc{ogcm} provides an 
average ice thickness, $h$, 
and surface concentration of the ice cover, $c$.

Let the wave energy density spectrum at point  $x$ and time $t$ be denoted $S(\omega;x,t)$, 
where $\omega$ is angular frequency.
The domain $x\in(0,L)$ is considered to be free of waves initially.
An incident wave spectrum is prescribed at the left-hand end of the domain, $x=0$. 
A Bretschneider incident wave spectrum, which is defined by a peak period and a significant wave height, is used for the examples considered here.
The incident spectrum is transported through the transect according to the 
energy balance equation
\begin{equation}\label{eqn:trans}
\frac{1}{c_{g}}
\frac{\partial S}{\partial t}
+
\frac{\partial S}{\partial x}
=
-\diss
S.
\end{equation}
The quantity $c_{g}$ is the group velocity of the waves.
\citet{Wiletal13b} provide numerical evidence that \textsc{miz} widths and floe size distributions predicted by the \textsc{wim} are insensitive to wave dispersion, 
\ie dependence of the group velocity on wave frequency. 
A constant value of $c_{g}$ is therefore used in the transport equation for the present application.
For simplicity, it is assumed that the wave spectrum travels the length of the transect in the time frame, \ie $c_{g}=L/T$.
The quantity $\diss=\diss(\omega;x,t)$ is the so-called attenuation coefficient, 
\ie the exponential attenuation rate of wave energy per meter.

The above version of the transport equation ignores nonlinear transfer 
of wave energy between frequencies. 
Wave energy input due to winds is also neglected.
Attenuation is assumed to be caused solely by the presence of ice cover.
Long-range attenuation of swell and attenuation due to wave breaking and whitecapping are ignored.   

The value of the attenuation coefficient, $\diss$, depends on the properties of the ice thickness and surface concentration.
It also depends on the floe size distribution, which is an output of the \textsc{wim}. 
The floe size distribution is defined by the probability of finding a floe with diameter greater than $\fdm$ in the vicinity of a point $x$ at time $t$, which is denoted $\pdt(\fdm;x,t)$.
The floe size distribution, and hence the attenuation coefficient, 
vary suddenly in time if wave-induced ice fracture occurs.
The occurrence of fracture, in turn, depends on the wave energy spectrum $S$.
Calculation of the wave spectrum, $S$, and the floe size distribution, $\pdt$, are therefore coupled via the attenuation coefficient.

The fracture model derived by \citet{Wiletal13a} is used.
The model is based on strains imposed on the ice cover by wave motion, in which the ice is modelled as a thin-elastic beam.
The fracture condition is
\begin{equation}\label{eqn:strcrit}
\str\geq \eps_{c}\sqrt{-2/\log(\mathbb{P}_{c})},
\end{equation}
where
$\str = 2\sqrt{m_{0}[\eps]}$
is the strain imposed on the ice by the waves. 
The quantity $m_{0}$ is the zeroth moment of strain. 
The $n$th moment of strain is defined as
\begin{equation}
m_{n}[\eps]
=
\frac{h}{2}
\int_{0}^{\infty}
\omega^{n}
\kappa^{2}(\omega)S(\omega)
\wrt \omega,
\end{equation}
where $\kappa$ is the ice-coupled wave number \citep{Wiletal13a}. 
The quantity $\eps_{c}= \sigma_{c}/Y^{\ast}$ is the fracture limit of the ice, 
where $\sigma_{c}=1.76\exp^{-5.88\sqrt{\nu_{b}}}$ is the flexural strength of the ice \citep{Tim&OB94},
and $Y^{\ast}=10(1-3.5\nu_{b})-1$ is the effective Young's modulus of the ice
\citep{Wiletal13a}, 
in which $\nu_{b}$ denotes brine volume fraction.
Finally, the quantity $\mathbb{P}_{c}$ is a chosen critical probability.
Here, as in \citet{Wiletal13a,Wiletal13b}, the limit for a narrow-band spectrum 
$\mathbb{P}_{c}=\exp^{-1}$ is used.
In general, the critical probability can not be calculated.

Fracture occurs if  inequality (\ref{eqn:strcrit}) 
is satisfied for the prevailing wave and ice conditions.
If fracture occurs the maximum floe diameter, $\dmx$,  
is set to be half of the dominant wavelength, $\lambda$. 
The dominant wavelength is calculated from the dispersion relation for the ice-covered ocean 
\citep[e.g.][]{Ben&Squ12a},
using the dominant wave period
\begin{equation}
\tau=2\pi\sqrt{\frac{m_{0}[\eta]}{m_{2}[\eta]}},
\quad
\text{where}
\quad
m_{n}[\eta]
=
\int_{0}^{\infty}
\omega^{n}
S(\omega) 
\wrt \omega
\end{equation}
is the $n$th spectral moment of the surface elevation $\eta$
\citep{WMO98,Wiletal13a}.
The maximum floe diameter parameterises the floe size distribution $\pdt$.
The floe size distribution is otherwise considered known.
For the specific examples considered later the value of $\dmx$ represents an absolute maximum.
However, this need not be the case, \eg  $\dmx$ may be a diameter that a certain proportion of floe diameters are less than. 

A truncated power law floe size distribution is used 
in the examples presented in the next section.
The probability function, $\pdt$, for the truncated power law is
\begin{equation}\label{eqn:probD}
\pdt(d)
=
\left\{
\begin{array}{c c}
\ds
(d/\dmn)^{\gamma}
&
\text{if }
d\in[\dmn,\dmx],
\\[10pt]
0 
& 
\text{if } d\notin[\dmn,\dmx],
\end{array}
\right.
\end{equation}
where $\dmn=20$\,m is a chosen minimum floe diameter 
and $\gamma=2+\log_{2}(0.9)$ is the chosen exponent \citep{Toyetal11,Dumetal11,Wiletal13a}.
The attenuation coefficient used in the examples is
$\diss=
c\dissn/\dav$,
where $\dav$ is the mean floe diameter.
The quantity $\dissn=\dissn(\omega,h)$ is the attenuation rate per floe, calculated using the model
of \citet{Ben&Squ12a}.


\section*{Numerical implementation}

\subsection*{The model of \citet{Wiletal13b}}

\citet{Wiletal13b} applied spatial, temporal and spectral discretisations to the \textsc{wim}.
For each time step the following algorithm is applied.
\begin{enumerate}
\item
\textit{Advection:}
The discretised wave spectrum is mapped to an intermediate spectrum by solving the energy balance equation (\ref{eqn:trans}) without attenuation, \ie $\diss=0$.
\item
\textit{Attenuation:}
The intermediate wave spectrum is attenuated according the properties of the ice it has just travelled through.
\item
\textit{Fracture:}
In each cell, fracture condition (\ref{eqn:strcrit}) is applied.
If the wave spectrum in a given cell is sufficient to cause fracture, 
the maximum floe diameter in the cell is set to half of the dominant wavelength. 
However, if the dominant wavelength is greater than the existing maximum diameter, no change is made. 
\end{enumerate}
\citet{Wiletal13b} set the initial maximum diameters to a large value $\sim 500$\,m
in cells that the maximum floe diameter is not otherwise initialised.


\subsection*{A new model without spatial or temporal discretisations}

The ice-cover acts as a low pass filter to the waves, \ie the attenuation rate decreases as frequency decreases/period increases.
Consequently, the wave spectrum becomes increasingly skewed towards low frequencies the further it propagates into the ice-covered ocean, and the associated wavelength becomes larger. 
The maximum floe diameter therefore increases in the \textsc{miz} with distance away from the ice edge.

The low-pass filter observation is the basis of an alternative numerical implementation of the 
\textsc{wim} proposed below, which does not require spatial or temporal discretisations.
The observation is translated to the assumption that the maximum floe diameter is due to ice fracture at the far end of the transect, $x=L$, or the furthest distance into the transect for which the wave spectrum is capable of causing fracture.

The problem therefore reduces to (i) determining if the wave spectrum at $x=L$ is sufficient to cause ice fracture, and, if not, (ii) determining the furthest point into the transect at which the wave spectrum is capable of causing ice fracture.
However, the wave spectrum and the maximum floe diameter must be calculated simultaneously, 
because, as noted in the previous section, they are coupled via the attenuation coefficient.

Suppose that a trial value is specified for the maximum floe diameter, and hence the 
attenuation coefficient is known.
The wave spectrum at $x=L$ is then obtained directly from the energy balance equation (\ref{eqn:trans}).
The associated strain, $\str_{L}(\dmx)$, and wavelength, $\lambda_{L}(\dmx)$, at $x=L$ are calculated as described at the end of the previous section.
Three scenarios are possible:
\begin{enumerate}
\item
the maximum floe diameter is less than half the wavelength;
\item
the maximum floe diameter is greater than half the wavelength; and
\item
the maximum floe diameter is equal to half the wavelength. 
\end{enumerate}
The wavelength and the maximum floe diameter balance one another in scenario 3,
\ie the maximum diameter defines a floe size distribution that attenuates the wave spectrum 
such that the wavelength at $x=L$ is consistent with the maximum diameter.
Scenario 3 is typically sought.
The only exception is if the trial value for the maximum floe diameter is an initial value.
In that case, in scenario 1 the wave spectrum causes no additional fracture, and the maximum floe diameter is not altered.

The maximum diameter that balances the wavelength at $x=L$ is obtained by finding the value of $\dmx$ that satisfies the equation 
\begin{equation}\label{eqn:balance}
\frac12 \lambda_{L}(\dmx)=\dmx
.
\end{equation}
However, for this value of $\dmx$, fracture only occurs if the strain imposed by the wave spectrum on the ice cover at $x=L$ satisfies the inequality 
\begin{equation}\label{eqn:frac2}
\str_{L}(\dmx)\geq \eps_{c}\sqrt{-2/\log(\mathbb{P}_{c})}.
\end{equation}

Equation (\ref{eqn:balance}) can be solved numerically using a package root finding algorithm, for example, MatLab's \texttt{fzero} function.
A check of the fracture condition (\ref{eqn:frac2}) 
is then performed for the calculated value of $\dmx$.

\begin{figure*}
\centering
 \includegraphics[scale=1]{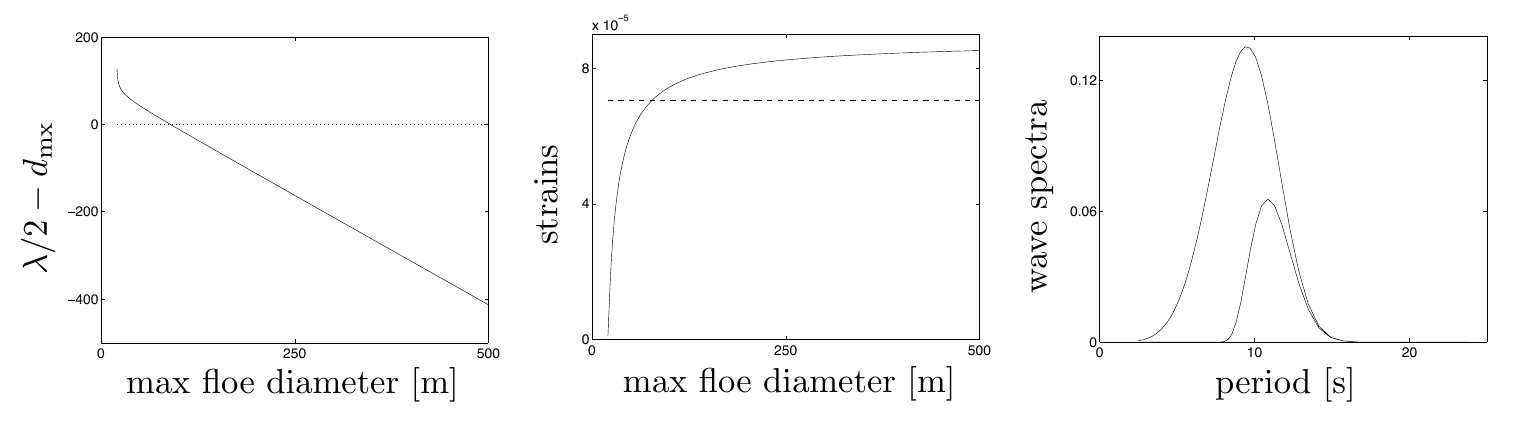}
\caption{\label{fig:Eg1} 
Results for an incident wave spectrum with 1\,m significant wave height and 9.5\,s peak period, and 
transect length $L=100$\,km.
Left-hand panel: half wavelength/maximum floe diameter balance function at $x=L$. 
Abscissa axis shown by dotted curve.
Middle panel: strain imposed on the ice by wave spectrum at $x=L$, $\str_{L}$, (solid curve), as a function of maximum floe diameter, $\dmx$. 
The strain threshold $\eps_{c}\sqrt{-2/\log(\mathbb{P}_{c})}$ is shown by the broken curve. 
Right-hand panel: incident wave spectrum and attenuated wave spectrum at $x=L$.
}
\end{figure*}

An example in which fracture occurs at $x=L$ is shown in figure~\ref{fig:Eg1}.
In this example, the incident wave spectrum has a 9.5\,s peak period  and 1\,m significant wave height.
The length of the transect is $L=100$\,km.
The left-hand panel shows that at $x=L$ the half wavelength and maximum floe diameter balance one another at approximately 89.5\,m.
The strains shown in the middle panel confirm that the wave spectrum is strong enough to fracture the ice cover when $\dmx\approx 89.5$\,m.
The incident wave spectrum and attenuated wave spectrum at $x=L$ are shown in the right-hand panel.
Skew of the attenuated spectrum towards large periods is clear.

For the example shown in figure~\ref{fig:Eg1} the maximum floe diameter is therefore set as $\dmx\approx 89.5$\,m. 
However, if the fracture condition is not satisfied at $x=L$, the problem is extended to search for the greatest distance into the transect that the wave spectrum is capable of causing fracture. 
The extension is necessary, as wave-induced fracture may be possible for a large proportion of the transect, despite not being possible for the entire transect.

Let $x=l$ be the greatest distance that fracture can occur.
The strain imposed on the ice cover by the wave spectrum at this point, $\str_{l}(\dmx)$, is such that
\begin{subequations}\label{eqns:ExtPrb}
\begin{equation}\label{eqn:ExtPrba}
\str_{l}(\dmx)= \eps_{c}\sqrt{-2/\log(\mathbb{P}_{c})}.
\end{equation}
Additionally, the wavelength at $x=l$, $\lambda_{l}$, must balance the maximum diameter, \ie
\begin{equation}\label{eqn:ExtPrbb}
\frac12
\lambda_{l}(\dmx)
=
\dmx.
\end{equation}
\end{subequations}
Equations (\ref{eqns:ExtPrb}a--b) are to be solved simultaneously for $l$ and $\dmx$.
In practice the equations are solved numerically as a set of one-dimensional equations:
\begin{enumerate}
\item[(i)]
by solving equation (\ref{eqn:ExtPrbb}) to give the maximum diameter that balances the wave spectrum, as a function of distance $l$, \ie $\dmx=\dmx(l)$; and
\item[(ii)]
substituting the expression for $\dmx=\dmx(l)$ into equation (\ref{eqn:ExtPrba}) and solving for the 
distance $l$.
\end{enumerate}

\begin{figure*}
 \centering
 \includegraphics[scale=1]{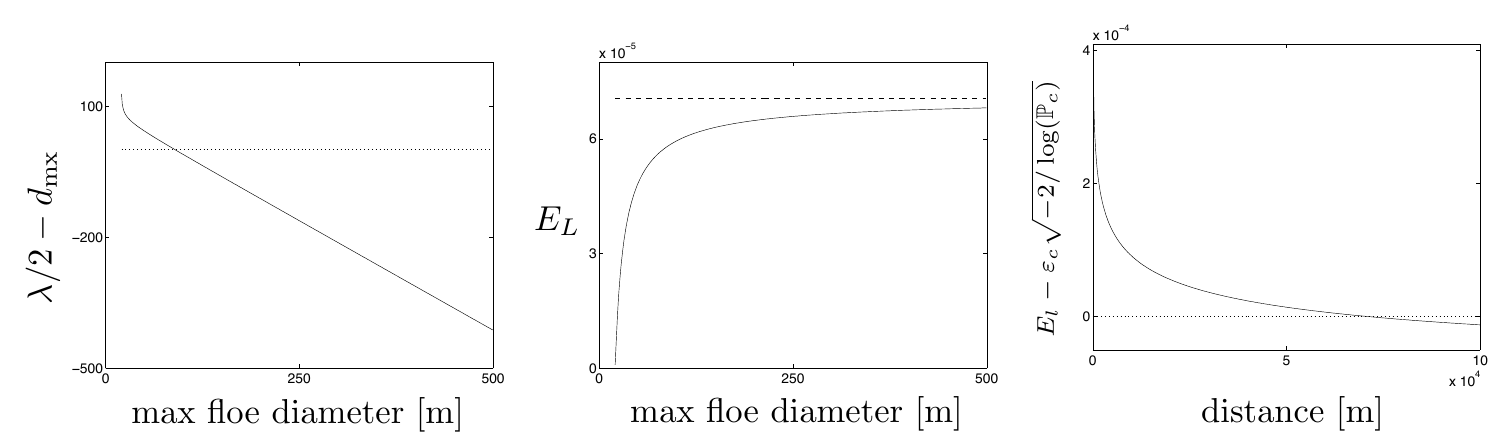}
 \caption{\label{fig:Eg2}
Results for an incident wave spectrum with 0.8\,m significant wave height and 9.5\,s peak period, and  transect length $L=100$\,km.
Left-hand panel: half wavelength/maximum floe diameter balance function at $x=L$. 
Middle panel: strain imposed on the ice by wave spectrum at $x=L$, $\str_{L}$, (solid curve), as a function of maximum floe diameter, $\dmx$. 
The strain threshold $\eps_{c}\sqrt{-2/\log(\mathbb{P}_{c})}$ is shown by the broken curve. 
Right-hand panel: the strain balance equation as a function of distance, $l$.
In left- and right-hand panels the abscissa axes are shown by dotted curves.
}
\end{figure*}

Figure~\ref{fig:Eg2} shows an example in which the extended method is necessary.
The example problem is identical to that considered in figure~\ref{fig:Eg1}, but using an 0.8\,m significant wave height for the incident spectrum.
As in figure~\ref{fig:Eg1}, the left-hand and middle panels show the balance equation and strains at $x=L$.
The half wavelength and maximum floe diameter again balance one another at approximately 89.5\,m.
However, the wave spectrum at $x=L=100$\,km is not strong enough to fracture the ice for $\dmx\approx 89.5$\,m (or for any $\dmx$ less than 500\,m).
The right-hand panel shows the strain balance equation as a function of distance $l$.
The largest distance at which the wave spectrum remains strong enough to fracture the ice is approximately 71.3\,km, \ie wave-induced fracture occurs over a substantial proportion of the transect.

The model proposed above differs from that of \citet{Wiletal13b} in more than the numerical implementation.
The new model uses a single maximum floe diameter for the transect, 
which, recall, represents a single cell in an \textsc{ogcm}. 
The floe size distribution, $\pdt$, for the entire transect is parameterised by the maximum floe diameter.
In contrast, the model of \citet{Wiletal13b} considers floe size distribution to be a local property. 
Wave-induced fracture is used to define a maximum floe diameter 
and hence floe size distribution $\pdt$ for each subcell of the transect.
The floe size distribution for the entire transect is generally not of the form $\pdt$. 

\begin{figure*}
\centering
\includegraphics[scale=1]{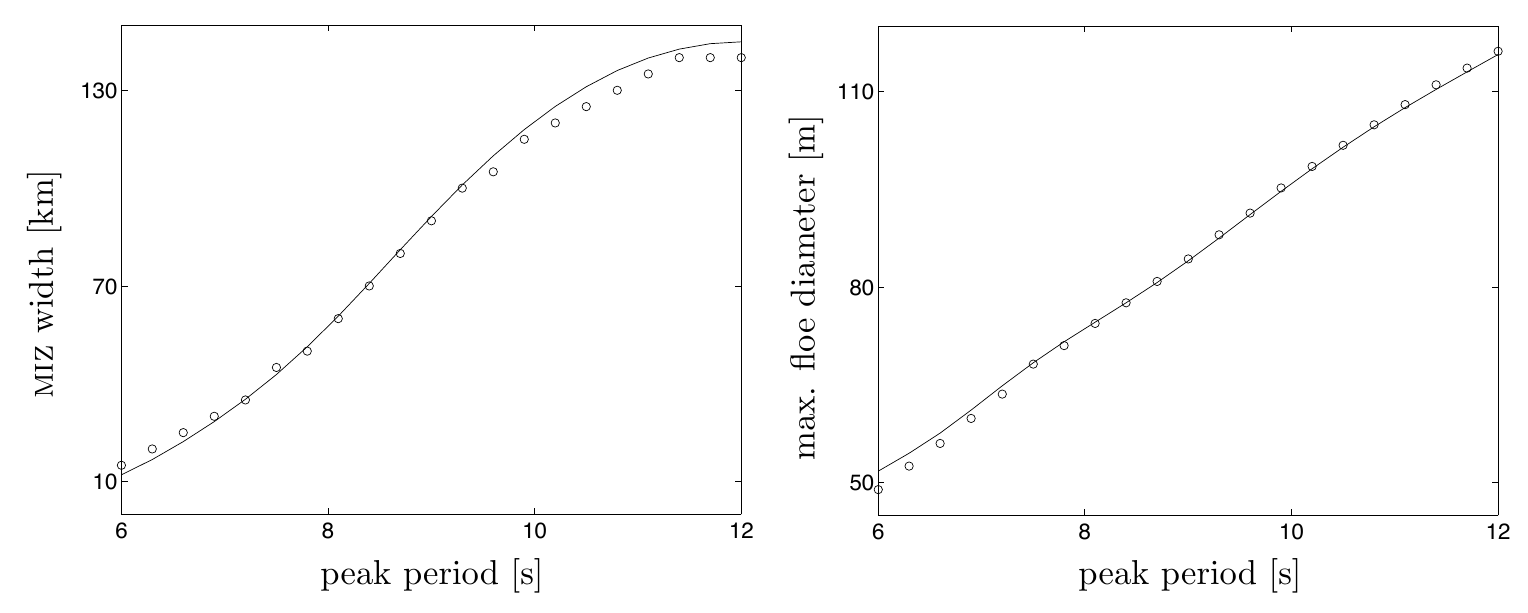}
\caption{\label{fig:Wiletal_comp} \textsc{Miz} width, $l$, and maximum floe diameter, $\dmx$, as functions of incident wave spectrum peak period.
The incident wave spectrum significant wave height is 1\,m.
Circles denote results produced by model of \citet{Wiletal13b}.
Curves denote results produced by model proposed in this work.
}
\end{figure*}

Figure~\ref{fig:Wiletal_comp} shows  results produced by the model proposed above and the model of \citet{Wiletal13b} for an example problem.
\textsc{Miz} width, defined as the distance of the transect over which wave-induced fracture occurs,
as a function of the peak period of the incident wave spectrum 
is shown in the left-hand panel.
Maximum floe diameter in the \textsc{miz}, as a function of peak period is shown in the right-hand panel.
The length of the transect is set to be very large so that the \textsc{miz} is not truncated.
The significant wave height of the incident spectrum is 1\,m.

\textsc{Miz} width and maximum floe diameter both increase as peak period increases.
The increase in \textsc{miz} width becomes less rapid as peak period increases.
The relationship between maximum floe diameter and peak period is approximately linear for the interval considered.

Given the different interpretations of the floe size distribution in  the two models, the agreement in the results is reassuring.
(Slight differences for the predicted \textsc{miz} widths can be attributed, in part, 
to discontinuities in the model of \citet{Wiletal13b} due to the discretisation of the transect 
--- a 5\,km cell length was used for the example.)
The results therefore indicate that the \textsc{wim} is insensitive to precise knowledge of the floe size distribution.


\section*{Summary and discussion}

A new version of the \textsc{wim} proposed by \citet{Dumetal11} and \citet{Wiletal13a,Wiletal13b} has been presented.
The model was based on the assumption that the maximum floe diameter 
occurs at the far end of the cell. 
At the far boundary the dominant wavelength is at its maximum, as the wave attenuation due to ice cover skews the wave spectrum towards large period waves.

The maximum floe diameter was used to define the floe size distribution in the cell.
A wavelength/maximum floe size equation was posed on the far boundary 
and solved to obtain the wave attenuation rate and floe size distribution, simultaneously.
In the case that the wave spectrum was not strong enough to fracture the ice at the far boundary, the problem was extended to search for the furthest distance into the cell that fracture occurs.

The new version of the \textsc{wim} was presented for an idealised transect geometry.
Example results for the width of the \textsc{miz} and maximum floe diameter due to wave-induced fracture, as functions of peak wave period were given.
The results were compared to results produced by the model of \citet{Wiletal13b}.
The two models gave almost identical results.
Close agreement was not guaranteed, due to the different interpretations of the floe size distribution in the two models.

Integration of the \textsc{wim} into \textsc{ogcm}s, \eg for climate studies, 
will require, in the first instance, an extension to two-dimensional geometries.
The key assumption of the model that the maximum floe diameter occurs at the far end of a cell can  be reinterpreted directly in the two-dimensional setting.
However, the validity assumption must be tested for directional wave spectra and multiple thickness categories.


\section*{Acknowledgements}

LB acknowledges funding support from the Australian Research Council (DE130101571) and the Australian Antarctic Science Grant Program (Project 4123). 
VS acknowledges funding support from the U.S. Office of Naval Research (N00014-131-0279).
Use of the Australian National Computing Infrastructure National Facility is provided by the 
National Computational Merit Allocation Scheme.


\end{document}